\documentstyle[12pt,epsf]{article}

\newcommand{\sect}[1]{\setcounter{equation}{0}\section{#1}\indent}

\begin{document}

\topmargin 0pt
\oddsidemargin 5mm
\def\bbox{{\,\lower0.9pt\vbox{\hrule \hbox{\vrule height 0.2 cm
\hskip 0.2 cm \vrule height 0.2 cm}\hrule}\,}}
%%%%%%%%%%%%%%%%%%%%%%%%%%%%%%%%%%%%%%%%%%%%%%%%
\def\a{\alpha}
\def\b{\beta}
\def\g{\gamma}
\def\G{\Gamma}
\def\d{\delta}
\def\D{\Delta}
\def\e{\epsilon}
\def\h{\hbar}
\def\ve{\varepsilon}
\def\z{\zeta}
\def\t{\theta}
\def\vt{\vartheta}
\def\r{\rho}
\def\vr{\varrho}
\def\k{\kappa}
\def\l{\lambda}
\def\L{\Lambda}
\def\m{\mu}
\def\n{\nu}
\def\o{\omega}
\def\O{\Omega}
\def\s{\sigma}
\def\vs{\varsigma}
\def\S{\Sigma}
\def\vphi{\varphi}
\def\av#1{\langle#1\rangle}
\def\pa{\partial}
\def\na{\nabla}
\def\hg{\hat g}
\def\un{\underline}
\def\ov{\overline}
\def\cF{{{\cal F}_2}}
\def\Hsl{H \hskip-8pt /}
\def\Fsl{F \hskip-6pt /}
\def\cFsl{\cF \hskip-5pt /}
\def\ksl{k \hskip-6pt /}
\def\pasl{\pa \hskip-6pt /}
\def\tr{{\rm tr}}
\def\tcF{{\tilde{{\cal F}_2}}}
\def\tg{{\tilde g}}
\def\shalf{\frac{1}{2}}
\def\nn{\nonumber \\}
\def\w{\wedge}
\def\tA{\tilde{A}}
\def\tF{\tilde{F}}
%%%%%%%%%%%%%%%%%%%%%%%%%%%

\def\cmp#1{{\it Comm. Math. Phys.} {\bf #1}}
\def\cqg#1{{\it Class. Quantum Grav.} {\bf #1}}
\def\pl#1{{\it Phys. Lett.} {\bf B#1}}
\def\prl#1{{\it Phys. Rev. Lett.} {\bf #1}}
\def\prd#1{{\it Phys. Rev.} {\bf D#1}}
\def\prr#1{{\it Phys. Rev.} {\bf #1}}
\def\prb#1{{\it Phys. Rev.} {\bf B#1}}
\def\np#1{{\it Nucl. Phys.} {\bf B#1}}
\def\ncim#1{{\it Nuovo Cimento} {\bf #1}}
\def\jmp#1{{\it J. Math. Phys.} {\bf #1}}
\def\aam#1{{\it Adv. Appl. Math.} {\bf #1}}
\def\mpl#1{{\it Mod. Phys. Lett.} {\bf A#1}}
\def\ijmp#1{{\it Int. J. Mod. Phys.} {\bf A#1}}
\def\prep#1{{\it Phys. Rep.} {\bf #1C}}

%%%%%%%%%%%%%%%%%%%%%%%%%%%%%
\begin{titlepage}
\setcounter{page}{0}

\begin{flushright}
IASSNS-HEP-98/32 \\
hep-th/9804019\\
April 1998
\end{flushright}

\vspace{5 mm}
\begin{center}
{\large Supergravity, the DBI Action and Black Hole Physics }
\vspace{10 mm}

{\large S. P. de Alwis\footnote{e-mail: dealwis@sns.ias.edu, 
dealwis@gopika.colorado.edu}~\footnote{On leave from 
Department of Physics, Box 390,
University of Colorado, Boulder, CO 80309. }}\\
{\em School of Natural Sciences, Institute for Advanced Study, 
Princeton NJ 08540}\\
\vspace{5 mm}
\end{center}
\vspace{10 mm}

\centerline{{\bf{Abstract}}}
 The assumptions behind the recently conjectured  
relation between gauge theory and supergravity are elaborated on.
It is pointed out that the scaling limit that preserves
supergravity solutions, gives the entire DBI action on the 
gauge theory side, but in the low energy limit the relation between 
the conformal field theory and Anti-de Sitter supergravity emerges. We 
also argue that recent work on these issues 
 may help in understanding
the physics of five (four) dimensional black hole with three (four)
charges in the so-called dilute gas region.

\end{titlepage}
\newpage
\renewcommand{\thefootnote}{\arabic{footnote}}
\setcounter{footnote}{0}

\setcounter{equation}{0}
\sect{Introduction}
Recent work has highlighted the correspondence between supergravity solutions
for black holes (branes) and their thermodynamic properties, with
microscopic properties of 
D-brane gauge theory (for recent reviews see \cite{dy}). One important
case that has been studied in some detail has been the absorption
of scalars from D3 branes \cite{ik} where it was shown that exact 
agreement exists between certain supergravity calculations and the gauge
theory  calculation. On the basis of these results 
 it has been conjectured \cite{jm}
 that there is a ``new type of duality"
between gauge theory and gravity. 
In \cite{gkp} and \cite{ew} this conjecture was given a precise
form in terms of an ansatz relating the supergravity action at a classical
solution to a certain boundary conformal field theory.
To understand this connection better it would be necessary
to formulate string theory
in Ramond-Ramond backgrounds. However so far it has not been possible to find 
a conformal field theory in such backgrounds. The best one can do is to
consider the operator formulation of string perturbation theory in which
one can construct vertex operators for the RR field strengths.

In order to find the relation to gauge theory we need 
to  allow the string world sheet to have boundaries which are attached to 
D-branes (see \cite{jp} for a review). The techniques for dealing with 
this situation were developed in
\cite{clny} (for D9-branes) and were adapted for use in the general D-brane
situation in \cite{li},\cite{cs},\cite{ds}.  These arguments may
be summarized as follows. Let the sum of vertex operators for the fluctuations
around a flat background be denoted by $\hat{\cal L}_I$. This operator contains
in addition to  NSNS fluctuations such as $h_{\mu\nu}(x)$,
RR fluctuations such as $F_3=dC_2$. The equations of motion follow 
from the physical state condition,
\begin{equation}\label{brst}
Q\hat{\cal L}_I|\O>=0.
\end{equation}
where the state vector above is the ground state corresponding to the spherical
world sheet. The above equation translates into linearized equations of motion 
and Bianchi identities such as 
\begin{equation}\label{}
\nabla^2 h_{\mu\nu}=0,~~d^*F=0,~~dF=0\ldots
\end{equation}
Now we need to allow for the possibility of the world sheet
opening up giving the disc topology. The corresponding boundary
state $|B>$ was constructed in the nine-brane case in \cite{clny} and for the
general D-brane in \cite{li}and \cite{ds}. While this state is BRST invariant
by itself, in order to consider it as a vertex operator on the sphere on the
same footing as the other terms  one needs  a propagator $\Pi$,
and the total state one should consider is
\begin{equation}\label{}
|\Psi >=\hat{\cal L}_I|\O>+\Pi |B>.
\end{equation}
The BRST invariance of this state ($Q|\Psi >=0$) then leads to modified 
field equations such as 
\begin{equation}\label{feqns}
\nabla^2 h=gT_{\{\mu\nu\}}\d^{9-p}(x_{\perp})[\det (1+{\cal F})]^{1/2},~~~
d^*F_5=i{g\over 2}*J,...
\end{equation}
 In the above ${\cal F}=F-B$ where $F$ is the gauge field strength of the
 open string gauge field and $B$ is the two form NSNS field,
 $T$ is a certain ($\cal F$ dependent) matrix in space time, and $J$ is 
 a delta functional current with support at the location of the D-brane.
 
These arguments admittedly yield only weak field equations in the weak string
coupling limit. However 
they have a unique non-linear generalization.
This follows from the fact that the only generally covariant action which
will yield the left hand sides of (\ref{feqns}) is the string
effective action (in our case for type IIA or IIB). Similarly the source
action turns out to be the DBI action.
One is thus led to the conclusion that the effective theory 
below the string scale is given by the sum of the supergravity (including
RR fields) action and the DBI action. i.e. 
\begin{equation}\label{actn}
S[\Phi ,l ]+S_{DBI}[\Phi_l;A],
\end{equation}
where $\Phi$ stands for the collection of supergravity fields and $\Phi_l$
is the boundary value of these fields at the location (denoted by $l$)
of the D-branes. By 
$A$ we mean both the gauge field on the brane and the transverse
components of the  10d gauge 
field which represent fluctuations in the location of the brane. To
make the discussion precise we will concentrate on the example of 
the D3 brane and its near horizon geometry\cite{jm}. In this case
the DBI action becomes the super Yang Mills action and the geometry
becomes that of $AdS_5\times S_5$, but we expect the arguments below to be
of more general validity. 

The geometry is given by the $AdS_5\times S_5$ metric 
(with Euclidean signature)
\begin{equation}\label{}
ds^2={R^2\over z^2}(dt^2+dx_i^2+dz^2)+R^2d\O_5^2.
\end{equation}
where $R=4\pi (gN)^{1\over 4}\sqrt{\a '}$. The boundary of this geometry  
may be taken to be a flat space at $z=0$ and a point (which for convenience
we will call the horizon) at $z=\infty $. 
Let us discuss quantum field theory in this background with the 
action (\ref{actn}). As in \cite{gkp} we will only consider fluctuations
that are constant on the $S_5$ so we effectively have a five dimensional
theory. The brane is  located at $z=l$ and is supposed to give an effective
description in the region $l>z\ge 0$ of the region ($z>l$) behind the 
``stretched horizon" at $z=l$ (see figure). The supergravity action 
in this region is,
\begin{equation}\label{sgactn}
S[\Phi ,l]=\int^l_0dz\int d^4x L[\Phi ].
\end{equation}
%%%%%%%%%%%%%%%%%%%%%%%%
\begin{figure}
\centering
\epsfxsize=2.5in
\hspace*{0in}\vspace*{.2in}
\epsffile{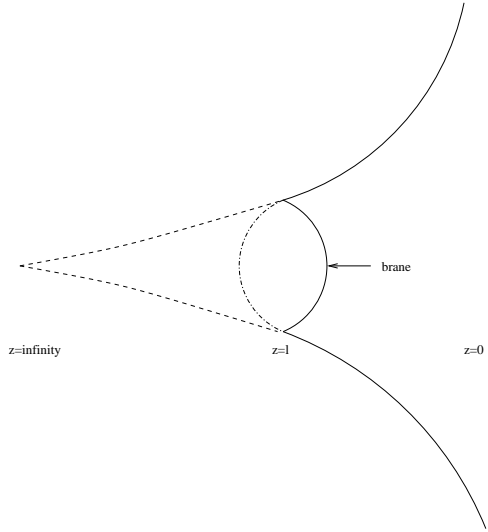}
\caption{The location of the brane in the AdeS space. The dashed region
 is replaced by the brane.}
\label{fig:Brane location}
\end{figure}
%%%%%%%%%%%%%%%%%%%%%%%%%
Now consider the quantum effective action (at its minimum) 
for fields which satisfy the
boundary condition $\Phi =\Phi_0$ at $z=0$ . (This quantum field
theory is effective  at scales below the string scale). 
\begin{equation}\label{}
e^{-\G[\Phi_{0}]}=\int [d\Phi ]_l|_{\Phi_{0}}[dA]
e^{-S[\Phi ,l]-S_{DBI}[\Phi_l,A,l]}.
\end{equation}
We have defined the measure above as $[d\Phi ]_l=\prod_0^ld\Phi (z)$.
Now the crucial assumption is that $\G$ is independent of $l$ the position
of the brane. It should be stressed that we do not have a justification for
this assumption beyond the fact that for minimally
coupled scalars the wave function is independent of l. In fact 
the situation may be
more complicated for  non-minimal fields \cite{ghkk}. 
While this does not necessarily make the
conjecture invalid it highlights the fact that it is 
far from being proven.
 
 Thus we may equate the right hand side evaluated at $l=\infty$
to its value at $l=0$. i.e.
\begin{equation}\label{}
\int [d\Phi ]|_{\Phi_{0}}[dA]e^{-S[\Phi ,\infty ]-
S_{DBI}[\Phi_{\infty}, A]}
=\int [dA]e^{-S_{DBI}[\Phi_{0}, A]}
\end{equation}
Where we have used the fact (see (\ref{sgactn})) that the supergravity
action vanishes for $l= 0$. At scales well below
the Planck scale we may approximate the supergravity functional integral
by its classical value $\Phi^c$ (which satisfies the same boundary conditions
as $\Phi$) 
so that we have the equation,
\begin{equation}\label{relation}
S[\Phi^c, \infty ]=\int^{\infty}_{0}dz\int d^4x L[\Phi^c ]=
W_{DBI}[\Phi_0] -W_{DBI}[\Phi_{\infty}].
\end{equation}
Here the generating functional for connected correlation functions in 
the gauge theory $W_{DBI}$ is defined
by 
\begin{equation}\label{}
W_{DBI}[\Phi_l]=-\log\int [dA]e^{-S_{DBI}[\Phi_l,A,l]}.
\end{equation}
Apart from an irrelevant constant (the second term on the right hand side
which is independent of $\Phi_0$) the equation (\ref{relation}) is the
ansatz of \cite{gkp} and \cite{ew}.
\sect{Large N limit of the DBI action}
Let us discuss now the precise conditions under which supergravity
is valid with the configuration of
the D-branes in question being again $N$ coincident D3-branes. The supergravity
solution for the ground state of the 3-brane theory is the extremal solution, 
\begin{equation}\label{ext}
ds^2=f^{-1/2}(-dt^2+\sum_{i=1}^{3}dx_i^2)+f^{1/2}(dr^2+r^2d\O_5^2).
\end{equation}
with
\begin{equation}\label{f}
f=1+{R^4\over r^4},~~R\equiv \sqrt{(4\pi gN)^{1/2}\a '}.
\end{equation}
 We also have for the Ramond-Ramond field
strength five-form $F_5\sim R\O_5$ where $\O_5$ is the volume of the
unit five-sphere and the dilaton is a constant.

The corresponding non-extremal solution for the metric, related to
some excited state (thermal state) of the brane theory, is 
\begin{equation}\label{nonext}
ds^2=f^{-1/2}(-(1-{r_0^4\over r^4})dt^2+\sum_{i=1}^{3}dx_i^2)+
f^{1/2}((1-{r_0^4\over r^4})^{-1}dr^2+r^2d\O_5^2).
\end{equation}
Here $r_0$ is the position of the horizon and is also a non-extremality
parameter. These solutions are valid up to the horizon provided that
the curvature at the horizon in string units is small, i.e.
$\a '{\cal R}\sim (gN)^{-1/2}<<1.$
 
 In the regime $r<<R$ the geometry of the extremal solution 
 is given by  dropping the ones in (\ref{ext}), (\ref{f}), i.e.
\begin{equation}\label{ads}
ds^2={r^2\over R^2}(-dt^2+\sum_{i=1}^{3}dx_i^2)+{R^2\over r^2}dr^2+R^2d\O_5^2.
\end{equation} 
 which is the by now well-known $AdS_5\times S_5$ geometry \cite{jm}.
 Similarly the near horizon geometry of the non-extremal solutions
 may be obtained by
 considering the region $r\sim r_0 <<R$. This gives
\begin{equation}\label{ads2}
ds^2={r^2\over R^2}(-(1-{r_0^4\over r^4})dt^2+\sum_{i=1}^{3}dx_i^2)+
{R^2\over r^2}(1-{r_0^4\over r^4})^{-1}dr^2+R^2d\O_5^2.
\end{equation}
 
 From these solutions it seems that the proper decoupling limit to consider is
 $g,\a '\rightarrow 0$ with $R$ fixed. We need to send $g\rightarrow 0$
 to suppress loop contributions, and  we are taking a large $N$
 limit with $gN = {R^2\over \a '}$  also becoming large to suppress the 
 corrections to supergravity. This is of course very similar to 
 the limit taken in \cite{jm} but here the scale which 
 defines the supergravity solution is held fixed (rather than going to
 zero as in \cite{jm}).

Now we need to  carry out  the  same ('t Hooft scaling) limit
 of the DBI action for D3-branes. As we will see below in this limit
all the terms  of this action are retained.

The DBI action for $N$ D3-branes (ignoring commutator terms, derivatives
of $F$ etc., and using a flat metric and ignoring the non-constant part of the
dilaton for simplicity ) is,
\begin{equation}\label{dbi}
I={1\over (2\pi)^3(\a ')^2g}\int\tr{(\det [\eta_{\mu\nu}+
(2\pi\a ')F_{\mu\nu}])^{1/2}} .
\end{equation}
In the above the trace is over $U(N)$, $g$ is the string coupling.
Also for simplicity of notation
we ignore the distinction between the gauge fields in the brane directions
and the transverse fields which are supposed to be
the moduli that are related to fluctuations in the positions of the 
branes. Let us expand the integrand schematically
in powers of $F^2$. Ignoring the first term (which is proportional to the
brane volume) we have,
\begin{equation}\label{}
I\sim{N\over gN}\int [\tr F^2+\a '^2\tr F^4+\ldots+
(\a ')^{2n-2}\tr F^{2n}+\ldots ].
\end{equation}
In the above $\tr F^4$ for example stands for a linear combination of
$\tr F^4$ and $(\tr F^2)^2$ and we have multiplied and divided the action
by $N$ to facilitate the passage to 't Hooft scaling. Note that the 
Yang-Mills coupling is given by $g_{YM}^2\sim g$. 

Now to get 't Hooft scaling we need to have just a factor of $N$ outside
the action as one takes the large $N$ limit. Thus we need to rescale the
gauge field by writing,
\begin{equation}\label{}
A=\sqrt{gN}\tA;~~F=\sqrt{gN}\tF;~~\tF=d\tA+\sqrt{gN}\tA^2.
\end{equation}
Then we have,
\begin{eqnarray}\label{}
I&\sim&N\int[\tr\tF^2+(\sqrt{gN}\a')^2\tr\tF^4+\ldots ]\nn
&\sim&N\int[\tr\tF^2+R^4\tr\tF^4+\ldots+R^{4n-4}\tr\tF^{2n}+\ldots ].  
\end{eqnarray}
where we have identified $R\sim (gN)^{1\over 4}\sqrt{\a '}$ as the 
scale of the dual supergravity solutions introduced in (\ref{f}). Now since
this is the scale which (in effect) should be fixed as one takes the 
limit $\a '\rightarrow 0$ all the terms in the
DBI action will contribute.

The above means that we have an infinite number of non-renormalizable terms
and the field theory description must break down for 
energies $E\ge O(R^{-1})$. The important point is that as long as  
$R^{-1}$ is much less than the Plank scale (which is becoming very large
in our limit) there is a regime $E\simeq R^{-1}$
in which higher order terms in the DBI action need to be considered on
the same footing as the classical supergravity action. In the infra-red
limit $E<<R^{-1}$ however the theory flows to a point on the
conformal fixed line defined by setting all but the $F^2$ coupling to zero
i.e. to the ${\cal N}=4$ super Yang-Mills theory. This large $N$ low energy
limit  is in fact equivalent to the `double scaling' limit of \cite{ik}.

Now in the DBI action, in addition to the gauge fields that we have written
down explicitly, there are also the moduli 
which come from the components of the
ten dimensional gauge field that are transverse to the D-brane. Let 
us call them $X$.
 A non-zero expectation value for these can be interpreted as the distance 
 between a probe brane and the rest
and should correspond to the distance $r$ in the supergravity
solution. This is the same sort of relation that one got in the matrix model.
Specifically we expect $r=R^2 X$. The  point is that the low energy limit 
in the gauge theory must correspond to the region $r\sim r_0 <<R$ in the 
supergravity solution.   
In this case one expects the 
correspondence described in \cite{jm} between the the conformal theory
on the brane (and its excited states) and the Anti de Sitter space supergravity
(\ref{ads}) (and its excited version (\ref{ads2})). Now as pointed out in
\cite{gkp},\cite{ew} the Kaluza Klein 
modes on the $S_5$ in (\ref{ads}) are of order
$R^{-1}$ and these masses agree with the dimensions of chiral operators in
the gauge theory. This agreement will of course be preserved by
 non-renormalization theorems and will not be affected by the DBI terms.
 On the other hand it is not clear in what sense all of string
 theory is kept since the string states have masses $O(gN)$ that are becoming
 large in the supergravity limit.

The above considerations will
 not affect the agreement of the low energy absorption 
 cross section calculated in the
weak coupling gauge theory with the supergravity calculation,
\cite{ik} since 
these are protected by  non-renormalization theorems, and will continue to
agree as one continues from strong to weak coupling keeping the energy
low. On the other hand the DBI terms will be seen in the intermediate 
region as pointed out by \cite{ghkk}. (Of course in (\ref{dbi}) above one
should now keep perturbations around the back ground metric as well as
the non-constant part of the dilaton). Our argument shows that even though
on the supergravity side only the  leading contribution is kept, on the gauge
theory side there are contributions (for  energies that are low compared
to the Planck scale but of the order of $R^{-1}$) coming from the higher order
terms in the DBI action.
\sect{Radiation from Black Holes in the Dilute Gas Approximation }
The ansatz of \cite{jm},\cite{gkp} and \cite{ew} 
is also relevant to the understanding
of radiation from brane configurations \footnote{See \cite{dy} for reviews.}. 
The main unresolved issue here (apart from the lack of understanding of
Schwarzchild black holes) is the fact that a weak coupling calculation
(from say an effective string theory) in flat space gives the correct 
radiation rate of the corresponding black hole \cite{dm},\cite{ms}.
 On the one hand there is
a calculation of an S-matrix in tree approximation
for the annihilation of modes living on the brane to produce a bulk field
which propagates out to infinity. This is a flat space 
calculation. On the other hand there is the supergravity calculation where
one calculates the absorption cross section for these bulk fields in the
near horizon geometry. The agreement of the entropy of the extremal
black holes calculated in these two ways (i.e. from the effective weakly
coupled string picture and the supergravity (Gibbons-Hawking) calculation)
was explained by arguing that in this case we are counting the degeneracy
of BPS states and therefore this should remain constant as we turn up 
the coupling. However in this non-extremal, non-BPS, situation 
this argument will clearly not work.
\footnote{Actually even in the BPS case it is still a puzzle why
the counting gives exactly the area entropy of Bekenstein and Hawking.}
 
 At this point we should discuss  the question of where the  brane is to be
 located. As we've argued earlier the physics is independent of this
 location but some locations may be more convenient than others
 depending on the particular questions one wishes to address.
  The standard picture is that the brane is a source for the
 supergravity solution that corresponds to it. One would then place it
 at the singularity (which it replaces) or at the horizon and think
 of it as replacing  the configuration behind the horizon. In this case
 one should not use a flat metric (and  constant dilaton etc)
 but one must consistently use the metric etc. coming from the supergravity
 solution at this point.\footnote{Precisely this observation was used recently
 to resolve an apparent paradox involving 
 `decoupled' NS five-brane theories \cite{sda}.}
 In this case the flat space calculations that are done in order to
 find the entropy and radiation rates are quite mysterious. One needs to
 make an argument as to why as one increases the coupling to go
 from flat space to a black hole configuration, the entropy and the rates
remain the same. However the picture
 that emerges from the discussion  in \cite{gkp},\cite{ew} is that 
 the brane (or conformal field theory) is at the boundary of the
 Anti de Sitter space and thus there is no need for any extrapolation.  
 It seems that this idea can be of relevance 
 for understanding the microscopic calculations of entropy and gray body
 factors for five and four dimensional black holes. Since it is 
 somewhat simpler we will concentrate on the five dimensional
 black hole but the 
 argument can be easily extended to the four dimensional case.
 
 The microscopic configuration (D-brane gauge theory)
 that gives the five dimensional black hole
 is one where (in type IIB) D-strings carrying momentum are bound to 
 D5-branes. The corresponding supergravity solution (with topology
 $R^2\times S^3\times S^1\times T^4$ 
 \cite{hs} is given (in the string frame) by
\begin{eqnarray}\label{metric}
ds^2&=&f_1^{-\shalf}f_5^{-\shalf}[-dt^2+dx_5^2+{r_0^2\over r^2}(\cosh \s dt
+\sinh\s dx_5)^2+f_1dx_idx^i] \nn
& & f_1^{\shalf}f_5^{\shalf}[(1-{r_0^2\over r^2})^{-1}dr^2+r^2d\O_3^2]\nn
e^{-2\phi}&=&g^2f_1^{-1}f_5\nn
H&=&2r_5^2\e_3+2r_1^2e^{-2\phi}*_6\e_3
\end{eqnarray}
In the above 
\begin{equation}\label{radii}
f_1=1+{r_1^2\over r^2},~~f_5=1+{r_5^2\over r^2},~~r_1^2={gQ_1\a '^3\over V},~
r_5^2=gQ_5\a ',~ r_0^2\sinh 2\s=2{g^2n\a '^4\over L^2 V}.
\end{equation}
where $V$ is the volume (divided by $(2\pi)^4$) of the four torus  around which
the five brane is wrapped, $Q_1,Q_5,n$ are integers giving the numbers of 
each type of brane and the momentum quantum (in units of $1\over L$) and
L is the radius of $S_1$.

Now as in the discussion after (\ref{ads}) the relevant decoupling limit
that should be taken here to  suppress closed string loop corrections
and higher order terms in the $\a '$ expansion is 
\begin{equation}\label{limit}
g\rightarrow 0,~ \a'\rightarrow 0;~ gQ_1,gQ_5\rightarrow\infty,
~~r_1,r_5,r_n\equiv r_0\sinh\s~
 fixed,~~V\sim\a '^2 .
\end{equation}

Now low energies in the D-brane gauge theory corresponds to the  region 
\begin{equation}\label{}
 r, r_0,r_n<<r_1,r_5.
\end{equation}
Note that this is equivalent to the so-called dilute gas region studied in
\cite{hs}. In this regime the metric is 
\begin{eqnarray}\label{}
ds^2&=&{r^2\over R^2}(-dt^2+dx_5^2)+{r_0^2\over R^2}(\cosh\s+\sinh\s dx_5)^2 
\nn
&+&{R^2\over r^2}(1-{r_0^2\over r^2})^{-1}dr^2+R^2d\O_3^2+{r_1\over r_5}dx_i^2
\end{eqnarray}
where $R^2=r_1r_5$.
The first three terms which give a space whose coordinates are $t,x_5, r$ is 
actually the metric of the BTZ black hole \cite{btz} in 3 dimensions.  
To see this\footnote{The relation of the {\it extremal} five dimensional black
hole to the extremal BTZ black hole was first observed in \cite{sh}. In
\cite{ss} an argument relating the non-extremal black hole to the BTZ
black hole was given. Our discussion is an adaptation of the latter.}
we make the coordinate transformation \cite{ss},
\begin{equation}\label{}
r^2+r^2\sinh^2\s=\r^2.
\end{equation}
Then we find that the metric becomes,
\begin{equation}\label{}
ds^2=ds^2_{BTZ}+R^2d\O_3^2+{r_1\over r_5}dx_i^2.
\end{equation}
where,
\begin{equation}\label{}
ds^2_{BTZ}=-{(\r^2-\r^2_+)(\r^2-\r^2_-)\over\r^2R^2}dt^2+
\r^2(d\phi-{J\over 2\r^2}dt)^2+{\r^2R^2\over \r^2-\r^2_+)(\r^2-\r^2_-}
d\r^2.
\end{equation}
where $M={\r_+^2+\r_-^2\over R^2}$ and $J={2\r_+r_-\over R}$ are the  mass 
and angular momentum of the BTZ black hole, $\r_+=r_0\cosh\s~r_-=r_0\sinh\s$,
and the angular coordinate\footnote{Note that while $R$ 
is not necessarily equal to
$L$ we would still require that it be integral multiple of $L$.}
 $\phi ={x_5\over R}$ 

It should also be noted that in this regime the dilaton is constant
$e^{-2\phi}=g^2{r_5^2\over r_1^2}$ and that there is a constant Ramond
Ramond flux.
\footnote{Note that the BTZ black hole is (locally) conformally flat. 
This follows from the fact that locally the metric is diffeomorphic
to AdS which is conformally flat.}

The near horizon back ground that we have obtained is actually an
exact solution of the string background field equations. This is 
because the BTZ black hole has been shown to be (an orbifold of) the
 SL(2,R) WZW theory \cite{hw} and
of course so are $S_3$ (which is a SU(2) WZW theory
 and $T_4$.\footnote{In fact in this case one can avoid the question of RR field
 backgrounds by taking the S-dual configuration of NS5-F1 and momentum.
 It is not clear that this is an advantage though since one still has
 a magnetic NS background.}The conclusion is that when the parameters
of the brane configuration are in the dilute gas region there is a
sigma model CFT description  in the near horizon region as well as 
in the asymptotically flat region. This means that one need not confine
the discussion to $gN>>1$ and perhaps this explains 
why the effective string calculations agree with
supergravity.

This brings us to the question of where the brane (CFT) should be 
located in relation to the supergravity solution. The  usual picture has
been to consider the string (D-brane) picture as a model for Hawking
radiation in flat space with the black hole being replaced by the collection of
D-branes. This is expected to be a weak coupling description with the
space being flat. The agreement of the entropy with the geometrical calculation
was in that case attributed to supersymmetry. The agreement beyond extremal 
BPS limit for radiation rates including grey body factors in the dilute
gas approximation remained a mystery. 
In the light of recent work 
\cite{ik}\cite{jm},
\cite{gkp}, \cite{ew} and the above considerations a somewhat different 
interpretation which does not rely on supersymmetry appears to emerge. The
point is that the geometry is fixed to be the black hole (or rather string)
metric even in the limit $g\rightarrow 0$ (with $r_1,r_5$ fixed). 
So while both closed string loops
and $\a '$ corrections are suppressed a non-trivial background is present.
Then one interprets the string action  as 
an effective boundary action which represents the black
hole configuration. It should be emphasized that  this effective 
string is in fact a closed string and the fact
that the open string coupling $gN$ is large is irrelevant to
the calculation.
 Of course we have made the
assumption that the functional integral over the gauge field even
in the non-abelian case reduces to a effective string action.

This picture is in accord with the recent observations
that the entropy of these black holes can be obtained from a 
boundary conformal field theory \cite{as}.
What we are suggesting here is that this picture may give also a natural
explanation as to why the emission rates and gray body factors 
agree with the supergravity calculation.
  
 Thus one may argue as follows. Far from the black hole
 $r>>R$ we have flat space. Near the black hole $r<<R$ we have 
 $AdS_3\times S_3\times T_4$. Now the gray body factor calculation \cite{ms}
 depended on  solving the Klein-Gordon equation in just these two regions
 and matching the solutions. Thus one could model the relevant physics
 by replacing the original geometry by $AdS_3\times S_3\times T_4$
 and a boundary with a flat metric as in the version of the ansatz for $AdS_5
 \times S_5$ discussed in\cite{ew}. Thus one expects the process
 taking place in the bulk to be mirrored by the conformal field theory
 on the boundary which is in this case an effective string theory. This
 picture is thus an alternative to the correspondence principle 
 \cite{hp}as a way of 
 explaining the agreement between the  microscopic calculations and
 the supergravity calculations of black hole thermodynamics.

On the flat space boundary one has the D5-D1-momentum
configuration and the relevant limit is given in (\ref{limit}).
The important 
 point is that by integrating over the gauge field we get
 an effective (closed) string in a manner analogous to the way in which
   effective (closed)  (p,q) strings were obtained in (\cite{cs},\cite{ds}).
 Thus we expect the boundary gauge theory  functional integral
 \begin{equation}\label{}
\int [dX][dA] e^{iS_{DBI}[\phi^0,A,X]}=\int [dX]e^{iI[\phi^0,X]}.
\end{equation}
where the action in the expression on the right hand side 
is an effective string action\footnote{A suggestion for deriving this from
first principles has been given in \cite{hw}.} 
which we may write schematically as
\begin{equation}\label{}
I\sim {T\over 2}\int \pa X^{\mu}\pa X^{\nu}(\eta_{\mu\nu}+\kappa h_{\mu\nu}).
\end{equation}
where $\k\sim g\a'^4$. The $h$ field is of course (the boundary value of)
an external field
coming from the supergravity solution (generically called $\Phi$ earlier)
and according to  the ansatz in \cite{gkp},
\cite{ew}, one is to regard this as a classical source. 
   But what we have above is precisely the  effective
 string used in \cite{dm} to calculate the Hawking radiation rate.
  Thus
 the ansatz of \cite{gkp},\cite{ew} appears to give an explanation 
 of the agreement between 
 the string calcualtion and the supergravity calculation. Presumably 
 such arguments can be extended to explain the agreement of the grey
 body factors \cite{ms} as well.
\sect{Discussion} 
To conclude let us summarize our results. We have shown that
the ansatz of \cite{gkp} and \cite{ew} is equivalent to the assumption
that the quantum effective action for the low energy field theory is
independent of the location of the brane.  Then we showed that the
scaling limit that keeps the full supergravity D-brane solutions corresponds
to having the full DBI action on the gauge theory side. One is thus led
to a relation between the
full DBI action with an effective cutoff 
$R^{-1}<M_p;~R\sim (gN)^{1/4}\sqrt{\a '}$ and supergravity, but in the low
energy limit one could argue for a correspondence between for instance  the
N=4 Yang-Mills theory and supergravity on $Ads_5\times S_5$. Next we argued 
that this ansatz provided justification for the effective string model
\cite{dm} calculations which  agreed
with absorption rates of (and emission rates from) certain
black holes. The argument provides a rationale for treating the effective
string as if it were living in flat space and for 
the non-renormalization of the rates.

\sect{Acknowledgements:} I wish to thank Igor Klebanov and Edward
 Witten for useful comments.
   I would also
like to thank  Edward Witten for hospitality at the Institute for
Advanced Study and  the Council on  Research and Creative Work
 of the University of Colorado
for the award of a Faculty Fellowship. This work is partially supported by
the Department of Energy contract No. DE-FG02-91-ER-40672.

%%%%%%%%%%%%%%%%%%%%%%%%%%%%%%%%%%%%%%%%%%

%%%%%%%%%%%%%%%%%%%%%%%%%%%%%%%%%%%%%%%%%%%%%%%%
\end{document}